\documentclass[prb,aps,twocolumn,amsmath,amssymb,floatfix,
superscriptaddress]{revtex4-1}
\usepackage[dvips]{graphics}
\usepackage{graphicx}
\usepackage{color}
\usepackage{blindtext}
\definecolor{dred}{rgb}{0.75,0,0}

\begin{document}

\title{
{Spin-selective Aharonov-Casher caging 
in a topological quantum network}
}

\author{Amrita Mukherjee}
\email{amritaphy92@gmail.com}
\affiliation{Department of Physics, University of Kalyani, Kalyani,
West Bengal-741 235, India}
\author{Rudolf A.\ R\"{o}mer}
\email{r.roemer@warwick.ac.uk}
\affiliation{Department of Physics, University of Warwick, Coventry CV4 7AL, United Kingdom}
\author{Arunava Chakrabarti}
\email{arunava.physics@presiuniv.ac.in}
\affiliation{Department of Physics, Presidency University, 86/1 College Street, Kolkata, West Bengal - 700073, India}

\begin{abstract}
A periodic network of connected rhombii, mimicking a spintronic device, is shown to exhibit an intriguing spin-selective extreme localization, when submerged in a uniform out-of plane electric field. The topological Aharonov-Casher phase acquired by a travelling spin is seen to induce a complete caging, triggered at a special strength of the spin-orbit coupling, for half-odd integer spins $s \ge n \hbar /2$, with $n$ odd, sparing the integer spins. The observation finds exciting experimental parallels in recent literature on caged, extreme localized modes in analogous photonic lattices. Our results are exact.

\end{abstract}

\maketitle

\paragraph*{Introduction}
Ultracold (UC) atomic gases, loaded in optical potential landscapes provide a platform where condensed matter systems can be simulated exploiting an unprecedented control over the system. \cite{KatherineWright2019ComingSuperconductors} In the recent past, this enabled the observation of Anderson localization (AL) of atomic matter waves through path breaking experiments~\cite{roati,deissler,lucioni} that spurred waves of activity after almost fifty years since the proposition of this famous disorder-induced, quantum interference-driven phenomenon.~\cite{anderson,thouless,abrahams,borland} Experimental realizations of bosonic and fermionic Mott insulators~\cite{greiner,jordens}, and studies on the spatial correlations and density fluctuations in bosonic and fermionic UC atomic systems~\cite{folling,rom} have widened the canvas. Simulations of spin dynamics and phase transition using UC atoms have generated the possibility of devising new electronic, or even atom-based devices~\cite{simon}. Experiments on a gas of $^{87}$Rb atoms~\cite{campbell-1}, a theory of exotic quantum phases in a spin-orbit (SO) coupled spin-one bosonic system~\cite{dassarma}, motivated by experiments on itinerant magnetism in SO coupled Bose gases~\cite{campbell-2}  or, experiments on a two-orbital fermionic quantum gas of $^{173}$Yb atoms~\cite{riegger} have provided an inspiring canopy of results that gets further illumination from recent experiments revealing the intricacies of SO coupling in UC atomic gases~\cite{lin,dalibard,wang,cheuk}.

SO coupled spin-$1/2$ particles have been studied quite recently, revealing rich physics~\cite{zhu,edmonds,zhou} in respect of the AL phenomenon. Comparatively, hardly any effort is exerted to the physics of particles with spin $> 1/2$ and including both fermionic and bosonic spin states in the context of spin polarized transport or localization aspects. But such studies demand attention, especially after their experimental realization in the UC atomic systems discussed above. This motivates us to undertake a study of the spectral properties of particles with spin $s \ge 1/2$ propagating in an infinite array of rhombii. Such a lattice was previously considered by Aharony et al.~\cite{aharony} for $s=1/2$ as a model of a periodic spintronic device.
\begin{figure}[tb]
\begin{center}
(a)\includegraphics[width=0.65\columnwidth]{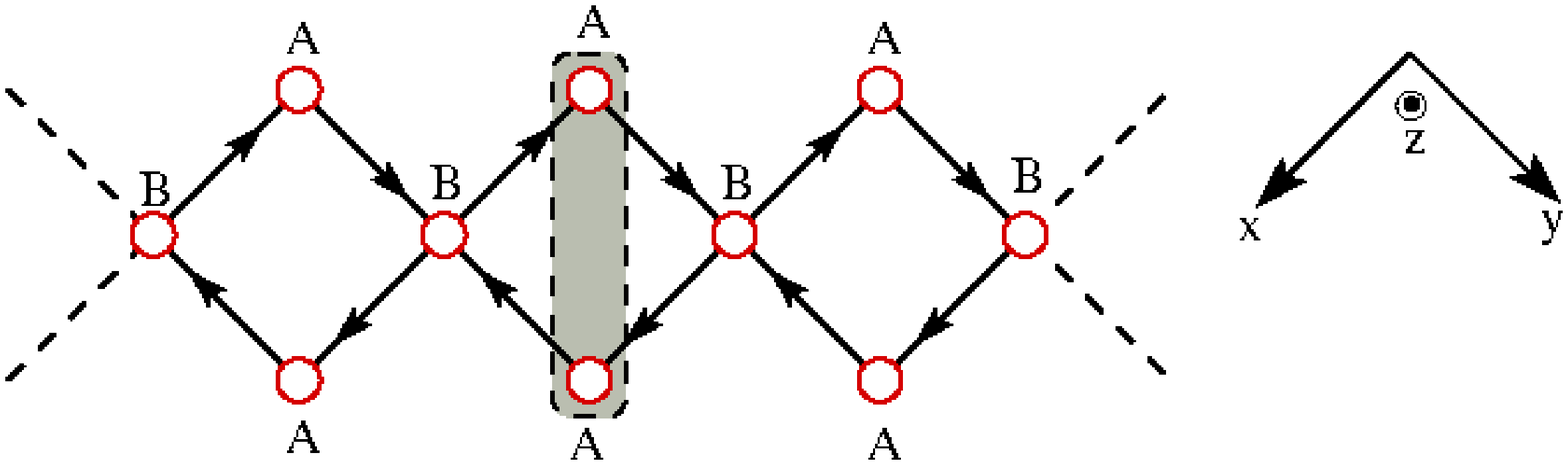}
(b)\includegraphics[width=0.50\columnwidth]{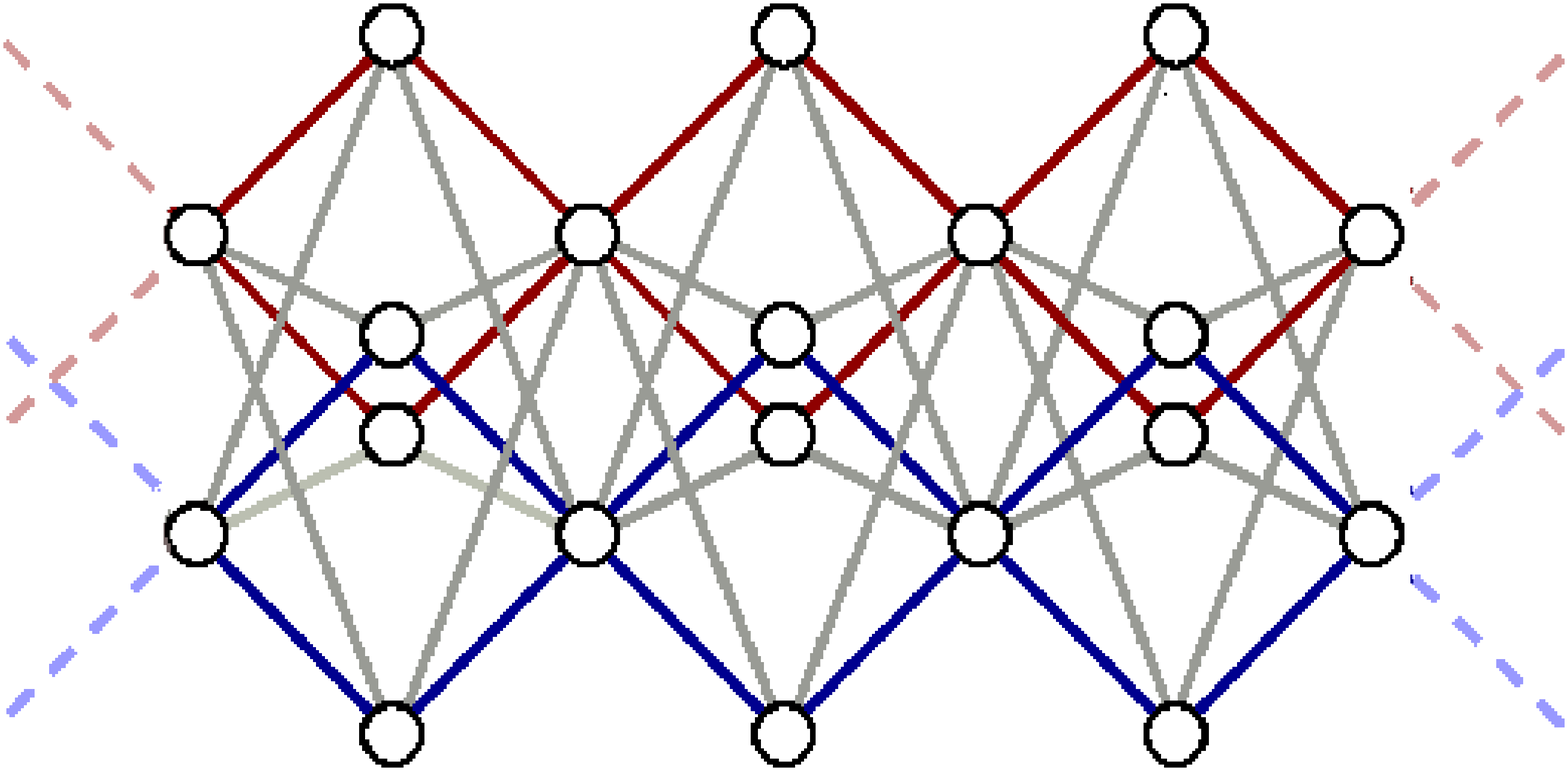}\\
\caption{(a) Schematic of an infinite rhombic chain in the $xy$ plane. The electric field is along the $z$ direction and the arrowheads depict the clockwise trajectory of the circulating particle from site ($\circ$) to site. The $A-A$ vertices entrapped in the shaded box and similar ones, have a pinned state at band centre (see text).
(b) Splitting of the rhombic geometry in `up' (brown) and `down' (blue) spin projections for spin $1/2$. The grey lines indicate the coupling between sites having opposite spin projections.
Dashed lines in (a) and (b) indicate the continuation to infinity. The coordinate system is shown in the top right corner.} 
\label{sample}
\end{center}
\end{figure}
We find a \emph{spin-selective} extreme localization effect that turns individual rombii into effective spin cages, reminiscent of  the well-known Aharonov-Bohm (AB) caging for electrons.\cite{julien-1}

We consider a, possibly neutral, particle with a non-zero magnetic moment and an arbitrary spin state $s \ge 1/2$ making an excursion in a periodic one dimensional network of rhombic tilings shown in Fig.~\ref{sample}(a). A similar effect was ushered into the domain of AL by Vidal et al.~\cite{julien-1,julien-2} for a large class of rhombic tiling threaded by a uniform AB magnetic flux~\cite{aharonov}. The competition between the periodicity of the potential landscape and the scale of area dictated by the magnetic field~\cite{julien-1} led to a rich spectrum characterized by a Hofstadter butterfly geometry.~\cite{hofstadter} Experiments confirmed and corroborated the AB caging effect through transport measurements in periodic superconducting wire networks~\cite{abilio} and even in normal metal networks,~\cite{naud} where the role of a half-flux quantum of magnetic flux has been verified. Very recently, photonic lattices, grown using ultrafast laser writing technology, and resembling exactly the rhombic array considered here and by Aharony et al.~\cite{aharony}, verified the AB caging effect~\cite{seba-1,seba-2,alex}, and witnessed the dramatic collapse of the entire optical spectrum into isolated sharp lines. A synthetic magnetic flux engineered such a collapse~\cite{seba-2} for light. Compact optical modes, exactly in the spirit of the compact localized {\it flat-band} states~\cite{sergej-1,sergej-2,sergej-3} in the electronic case were also observed. These observations open up a new possible direction towards topological photonics.
\begin{figure}[tb]
\begin{center}
(a)\includegraphics[width=0.42\columnwidth]{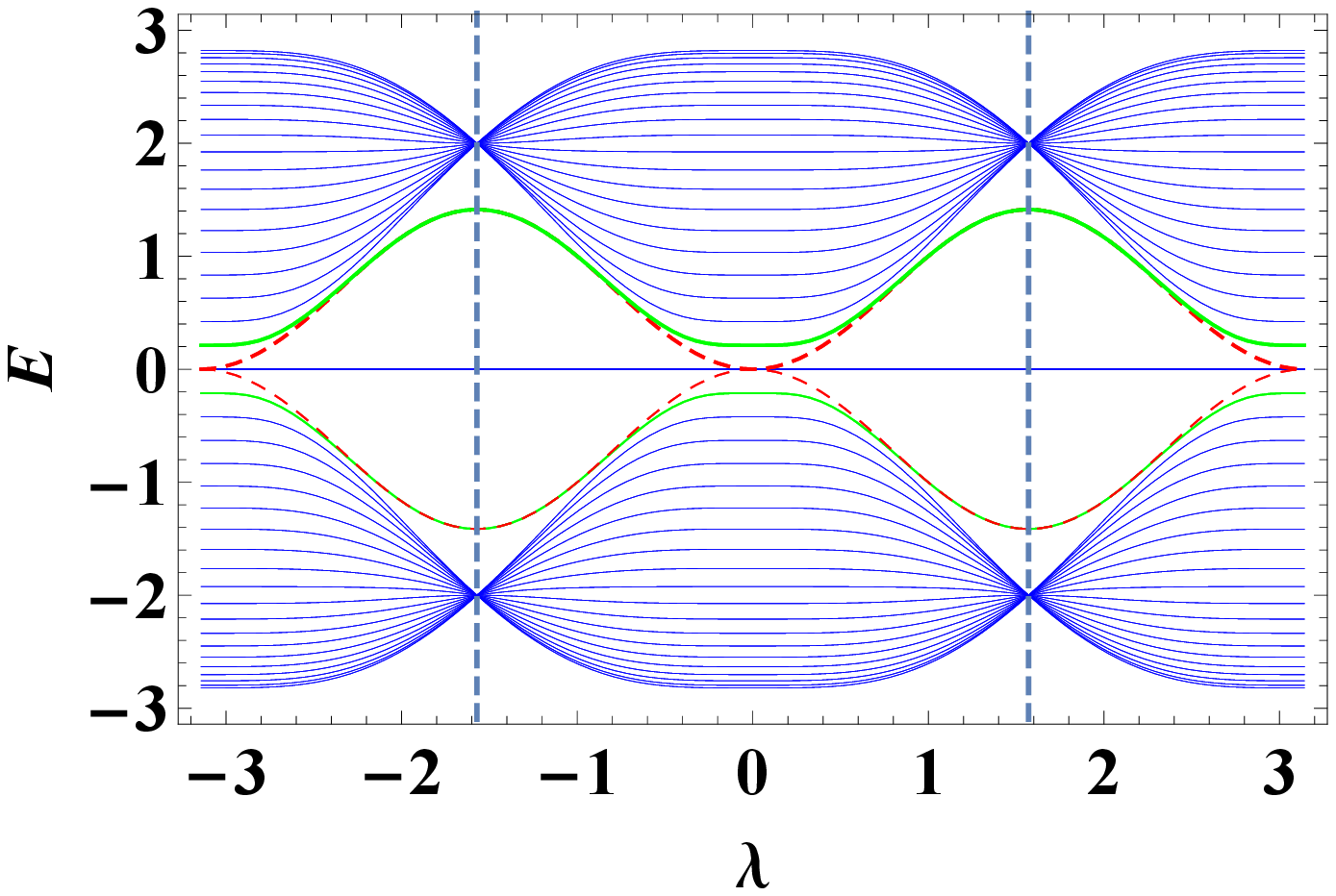}
(b)\includegraphics[width=0.42\columnwidth]{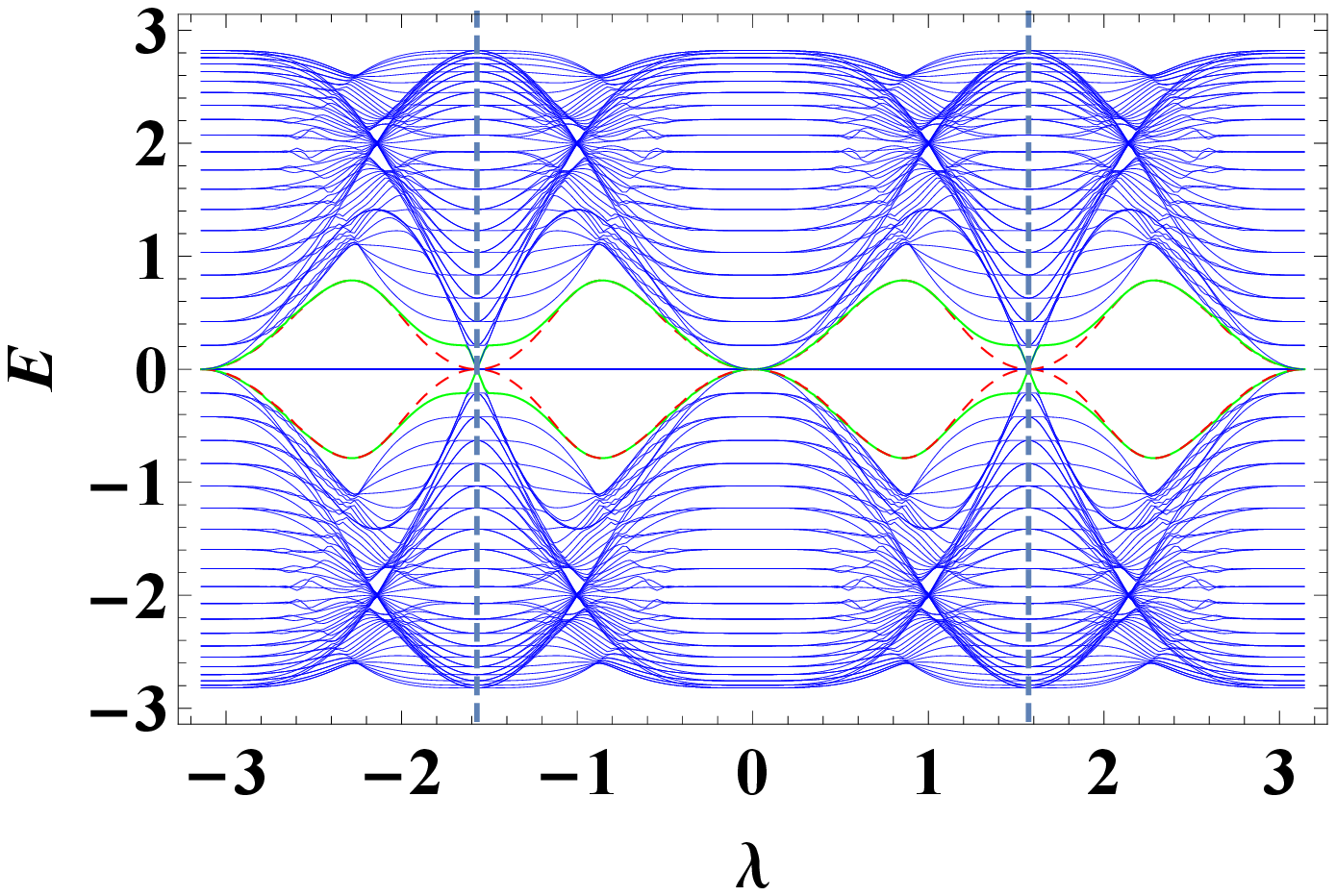}\\
(c)\includegraphics[width=0.42\columnwidth]{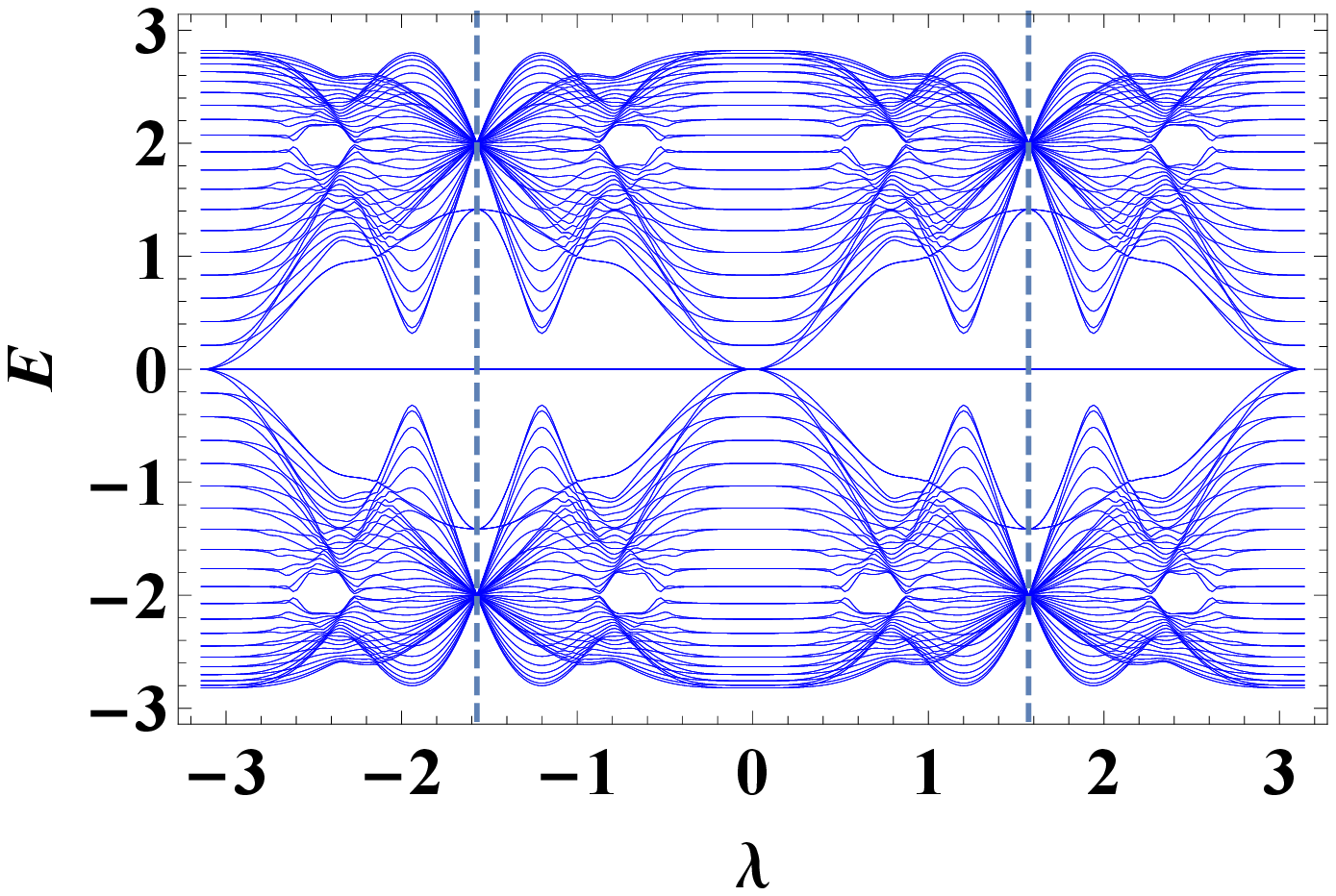}
(d)\includegraphics[width=0.42\columnwidth]{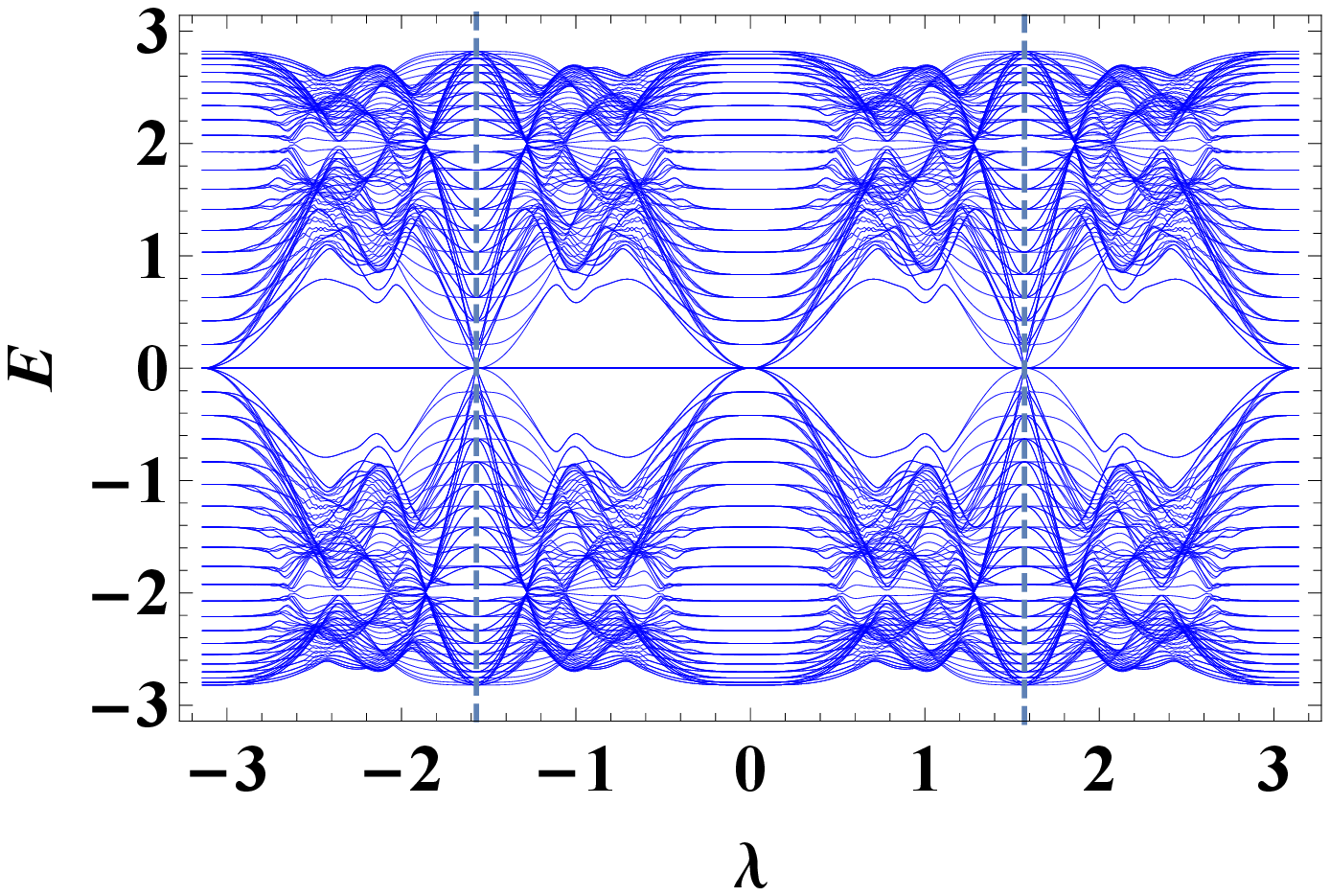}
\caption{
Variation of eigenstates against the SO coupling $\lambda$ (in units of $t$) for a $20$-loop rhombic array with $\epsilon=0$ and $t=1$ and spin (a) $s=1/2$, (b) $1$, (c) $3/2$, and (d) $2$. A flat, $\lambda$-independent band is seen at $E=0$ for all spins. The vertical dashed lines highlight $\lambda=\pm\pi/2$, while the dashed red and solid green lines indicate the topological edge states (see text).}
\label{dispersion}
\end{center}
\end{figure}

These results motivate us to use an out-of-plane electric field that leads to a non-abelian vector potential coupled to the spin of the particle, and results in a completely topological interference effect, as was initiated by Aharonov and Casher (AC)~\cite{casher}. The electric field introduces a Rashba-type spin-orbit (SO) interaction in the Hamiltonian of the system. 
We find that in an array of rhombii, threaded by a uniform electric field, the  AC effect~\cite{casher} results in an {\it extreme localization} of fermionic states with spin half-odd integer $s$, when the Rashba SO strength is tuned to a special value, while the bosonic, integer spin, counterparts never show any extreme localization. Still, for every spin, fermionic or bosonic, an SO strength-independent, {\it flat} band appears at the centre of the spectrum, along with spin projected topological edge states showing up at special SO coupling strengths. Remarkably, the collapsed, extreme-localized eigenstates for half-odd integer spins appear exactly at the energy eigenvalues as observed in a photonic AB-cage experiment by Kremer et al.~\cite{alex}

\paragraph*{The theory and the results}

A magnetic moment $\vec{\mu}$ moving with a velocity $\vec{v}$ in an electric field $\vec{\mathcal{E}}$ experiences a magnetic field in its own frame of reference, that couples with its spin. The resulting topological phase of the wave function is modelled via 
an AC phase factor in the hopping integral for $x' \rightarrow x$ of the tight binding Hamiltonian as $\exp[-i (\mu/\hbar c) \oint \vec{A}_\mathrm{SO} \cdot d\vec{r}]$. The vector potential $\vec{A}_\mathrm{SO} = (\hbar/4mc)\ \vec{\sigma} \times \vec{\mathcal{E}}$ involves non-Abelian matrices~\cite{oreg}. This is the SU$(2)$ analog of the U$(1)$ phase factor encountered in the AB effect. For convenience, we recast the phase factor as $\exp [i \lambda (\hat{n}\cdot\vec{\sigma}_s)]$ with $\lambda=\mu \mathcal{E}a/4mc^2$ representing the strength of the SO coupling, and $\vec{\sigma}_s=\{\sigma^{(s)}_x,\sigma^{(s)}_y,\sigma^{(s)}_z\}$ denoting the Pauli matrices of a spin $s$ particle.\footnote{We shall drop the correct but cumbersome $^{(s)}$ notation in the following and let $s$ be implied by the context of the discussion.} Here, $m$ is the mass of the circulating particle, $a$ in the length of each side of a rhombus, and $\hat{n}$ provides the direction of the effective magnetic field~\cite{avishai-1,matityahu}. 

%
The tight-binding Hamiltonian used here is given by,
\begin{equation}
\mathbf{\mathcal H}=
\sum_{x} \mathbf{c}^{\dag}_{x} \mathbf{\epsilon}_x  \mathbf{c}_{x} 
+ 
\sum_{\langle x,x' \rangle} 
\left[
\mathbf{c}^{\dag}_{x}  ~
\mathbf{t}_{xx'} e^{i\lambda \hat{n}\cdot\vec{\sigma}_s}
\mathbf{c}_{x'} 
+ 
\mathrm{h.c.}\right] .
\label{ham}
\end{equation}
%
The operators, $\mathbf{c}^\dagger_x$ and $\mathbf{c}_x$ are $2s+1$-component vectors while the on-site potential $\mathbf{\epsilon}_x$ and the hopping integral $\mathbf{t}_{xx'}$ are $(2s+1) \times (2s+1)$ matrices. The external electric field is chosen as $\vec{\mathcal{E}} = (0,0,\mathcal{E})$.
The general form of the exponential $e^{i \lambda \left(\hat{n}\cdot\vec{\sigma}_s\right)}$ for arbitrary spin $s$, is given in detail by Curtright et al.\cite{curtright} For example, when $s=1/2$ the explicit expression is 
$e^{i \lambda \left(\hat{n}\cdot\vec{\sigma}_{1/2}\right)} = \openone_{2} \cos \lambda + i\left(\hat{n}\cdot\vec{\sigma}_{1/2}\right) \sin \lambda$, 
while for $s=1$, we have 
$e^{i \lambda \left(\hat{n}\cdot\vec{\sigma}_{1}\right)} = \openone_{3} + i\left(\hat{n}\cdot\vec{\sigma}_{1}\right) \sin \lambda + \left(\hat{n}\cdot\vec{\sigma}_1\right)^2 \left(\cos\lambda - \openone_3 \right)$.
Here, $\openone_m$ denotes the $m\times m$ identity matrix.
The Schr\"{o}dinger equation $H\Psi=E\Psi$ can be cast as a set of difference equations
\begin{equation}
    (E \openone_{2m_s+1}  - \mathbf{\epsilon}_x) \mathbf{\psi}_{x,m_s} = \sum_{x'}
    \mathbf{t}_{xx'} e^{i\mathbf\theta_{xx'}(\lambda)}~\mathbf{\psi}_{x',m_s}
    \label{difference}
\end{equation}
where $\mathbf{\psi}_{x,m_s}$ is the appropriate spinor with $m_s=-s, -s+1, \ldots, s$. The $(2s+1) \times (2s+1)$ matrix $\theta_{xx'}(\lambda)\equiv \lambda \hat{n}\cdot\vec{\sigma}_s$ is the SO coupling dependent non-abelian AC-phase acquired, with appropriate sign, due a traversal from a site $x$ to its nearest neighbor $x'$.
Eq.~\eqref{difference} immediately reveals that the quantum network in Fig.~\ref{sample} becomes a $2s+1$ dimensional geometrical object in a spin-projected hyperspace in respect of the incoming spin $s$. Fig.~\ref{sample}(b) exemplifies this for $s=1/2$ where the linear network virtually `splits' into two, corresponding to the spin projections $m_s=\pm 1/2$ with `inter-spin' couplings dictated by the SO Hamiltonian. 

\begin{figure}[tb]
\begin{center}
(a)\includegraphics[width=0.4\columnwidth]{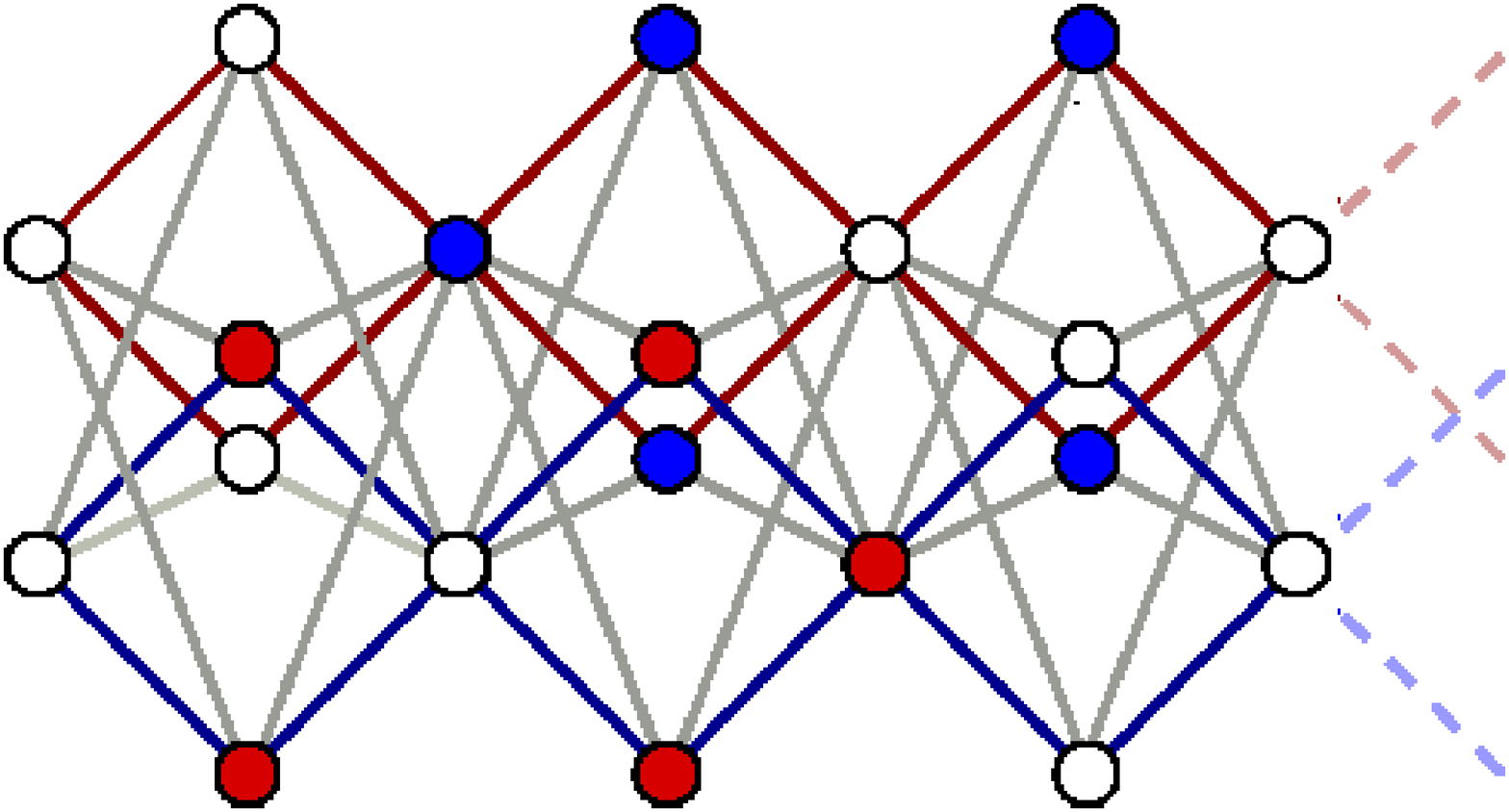}
(b)\includegraphics[width=0.4\columnwidth]{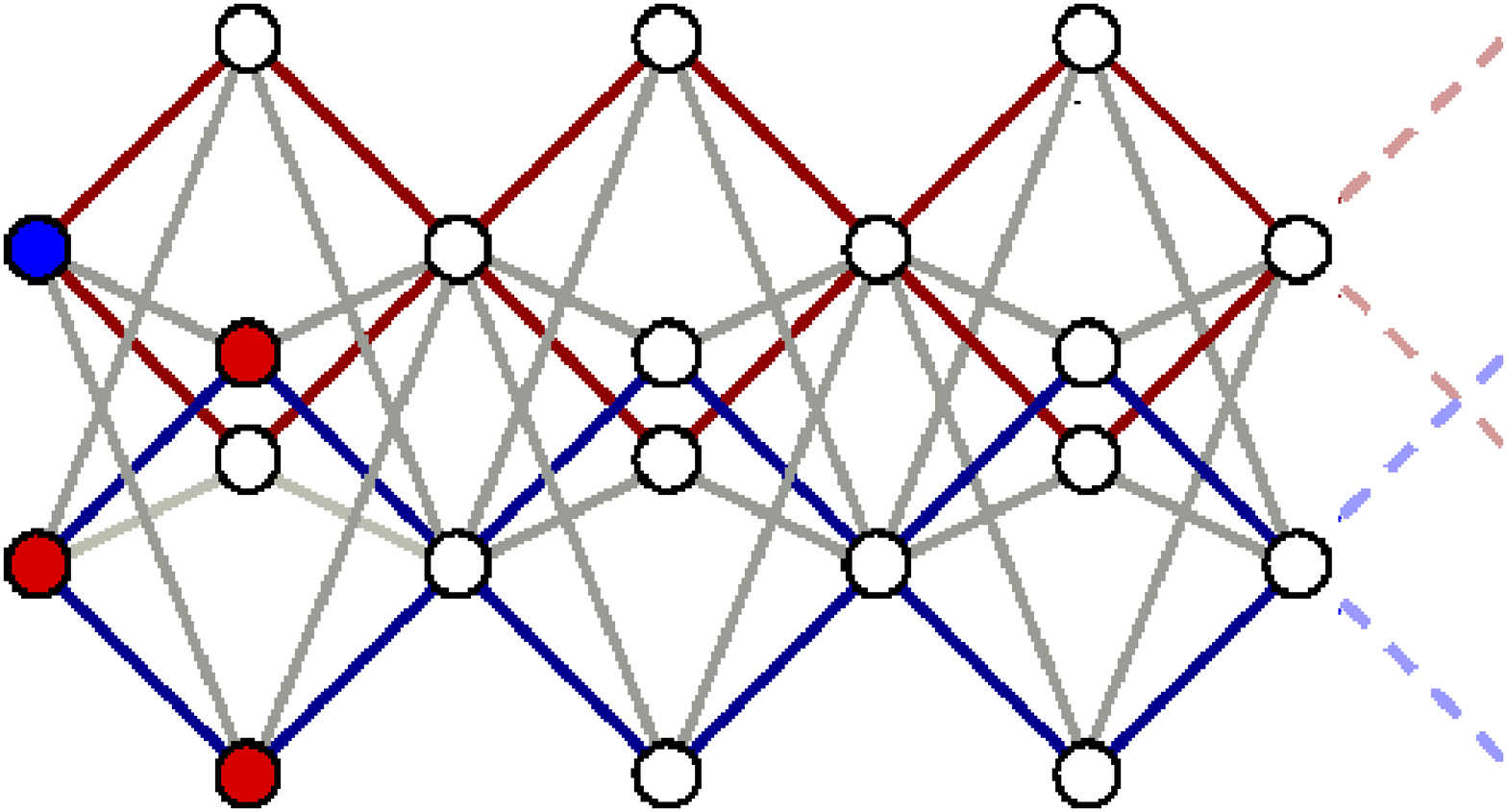}
\caption{Caging of amplitudes for $s=1/2$ with (a) $E=2$ and (b) $E=\sqrt{2}$ (one of the edge states) at $\lambda=\pi/2$. Blue and red solid circles indicate non-zero amplitudes in the $m_s=\pm 1/2$ 
projected spaces, respectively, while the open circles indicate sites $x$ where $\psi_{x,\pm 1/2}=0$. Other parameters are the same as in previous figures. } 
\label{ampli-dist}
\end{center}
\end{figure}

We have evaluated the energy spectra for spins $s=1/2, 1, 3/2$ and $2$ for a system of $20$ rhombii by exactly diagonalizing the Hamiltonian matrix with hard wall boundary conditions. The results are displayed in Fig.~\ref{dispersion} as functions of the SO coupling $\lambda$. For a general spin state $s$, the corresponding spin matrix is defined as $S = (\hbar/2) \sigma_s$. 
This `scaling' of the spin matrix $S$ puts all spins on the same footing, and becomes convenient as it makes the periodicity of the $E-\lambda$ variations the same for every spin --- much in the spirit of `zone folding' in dealing with energy bands in a crystal. 
Two features are of immediate importance: 
$(i)$ for every spin $s$, a flat, $\lambda$-independent band appears at $E=0$. Considering that $\lambda$ mimics the role of reduced momentum $k$ in a periodic potential, this state hence displays a \emph{non-dispersive} character. It is a localized state for which the amplitudes are \emph{pinned} at the top and bottom vertices ($A$) of each rhombus (caged in the shaded band) in Fig.~\ref{sample}(a).
The pinned profile can be verified by explicitly evaluating the amplitudes $\psi_{x,m_s}$ at every vertex $x=A$,$B$ of a rhombus for spin projection $m_s$, using Eq.~\eqref{difference}. It is easy to verify that, for $E=0$, the $\psi_{x,m_s}=0$, $x \forall B$, and for all spins, irrespective of the value of $\lambda$. This ensures localization, and an {\it extreme} one, as we shall come across later again. 
$(ii)$ For $s=n\hbar/2$ with $n$ odd, at special value of the SO coupling $\lambda=\pi/2$, the absolutely continuous subbands touch each other at just one point when $E=\pm 2$. This is displayed in Fig.~\ref{dispersion}(a) and (c) for spins $1/2$ and $3/2$. At $E=\pm \sqrt{2}$ we find two other isolated eigenvalues for all half-odd integer spins tested here.
\begin{figure}[tb]
(a)\includegraphics[width=0.4\columnwidth]{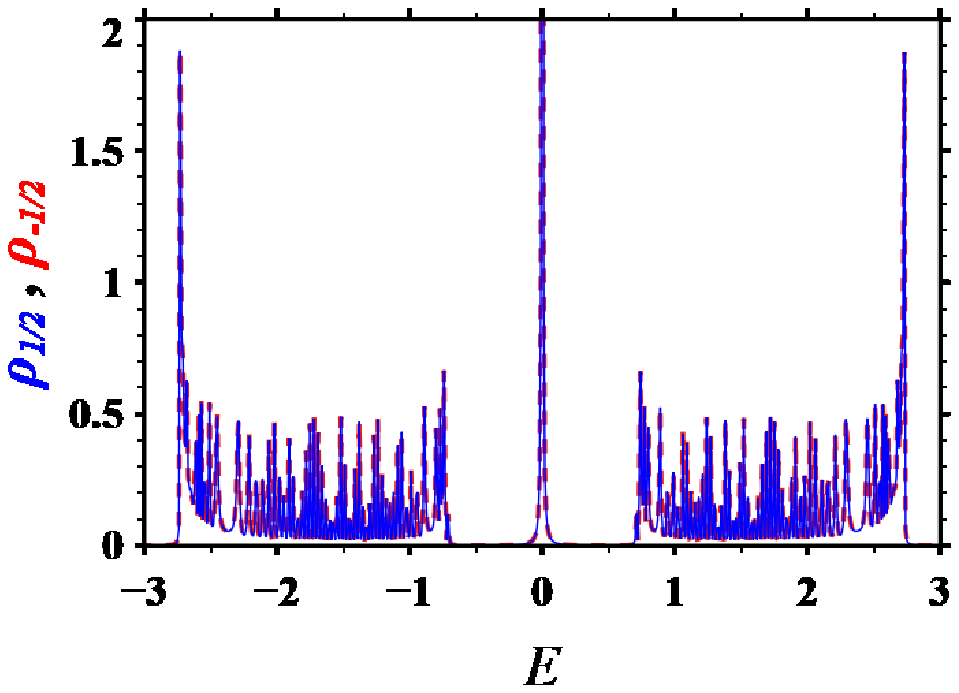}
(c)\includegraphics[width=0.4\columnwidth]{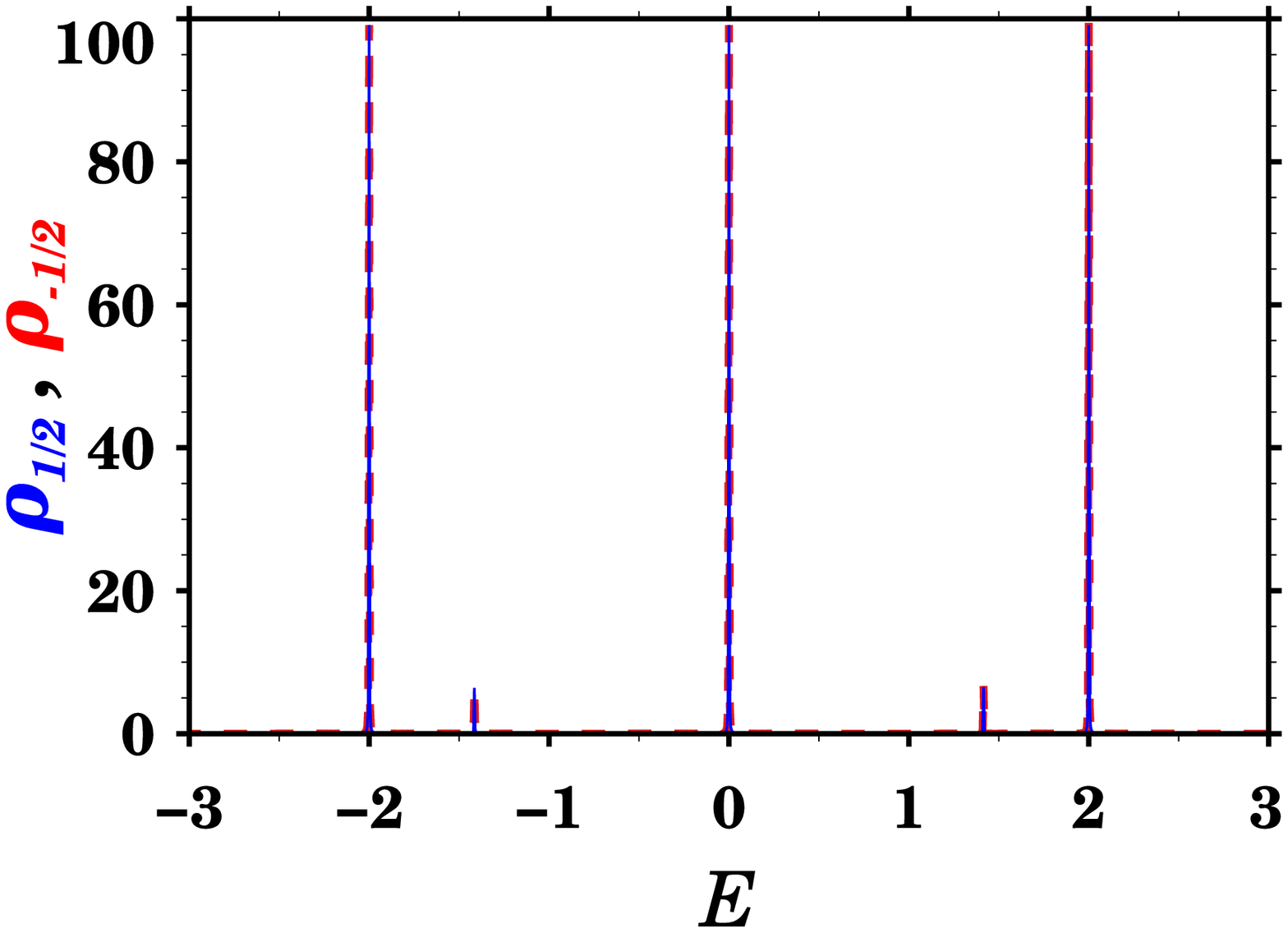}
\\
(b)\includegraphics[width=0.4\columnwidth]{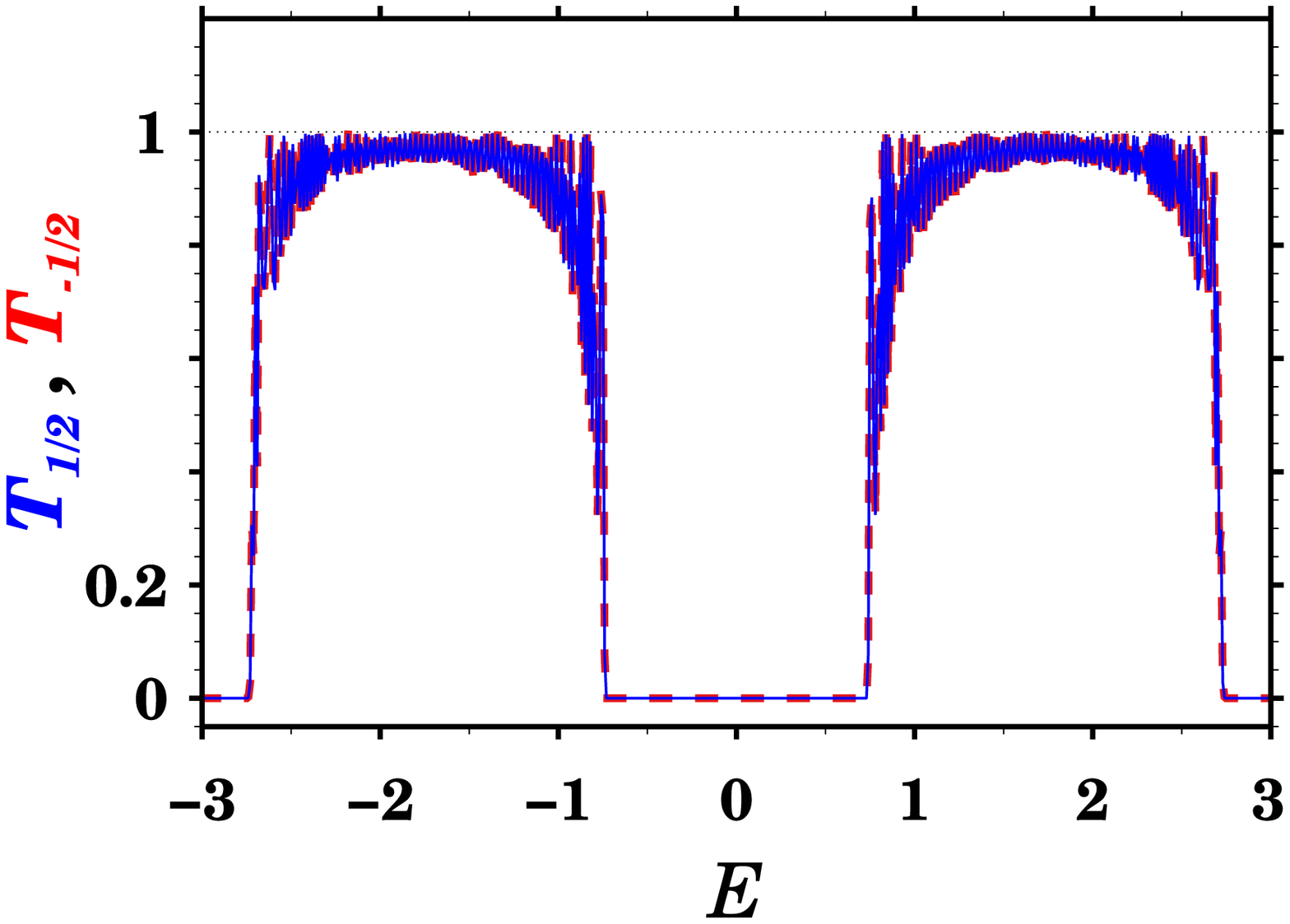}
(d)\includegraphics[width=0.4\columnwidth]{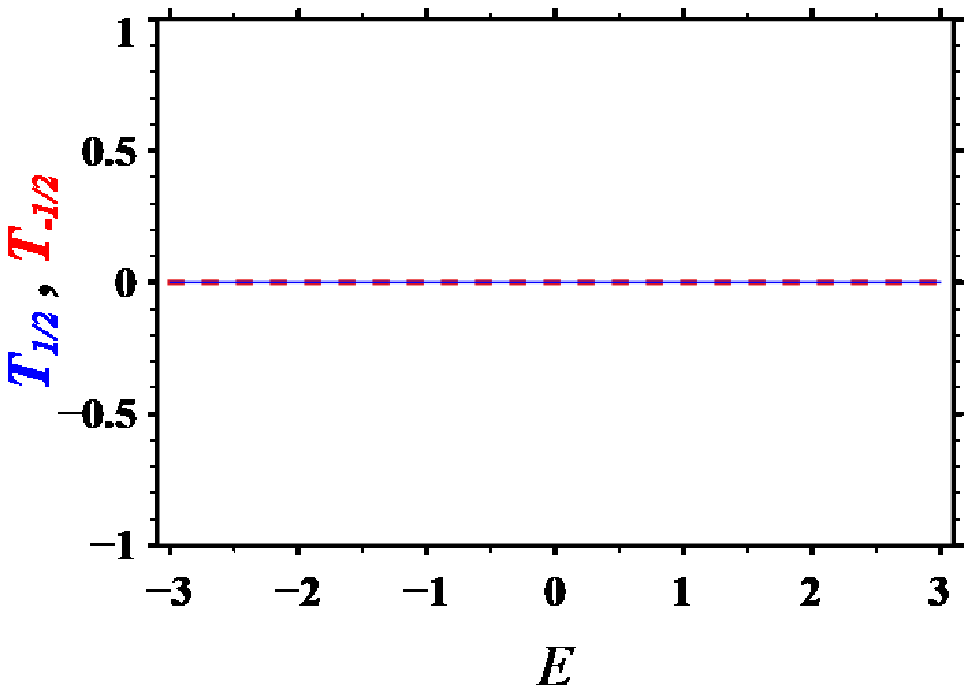}
\caption{
Plot of (a,c) DOS for an array of $30$ rhombii with hard wall boundary condition for an incoming spin with $s=1/2$, and (b,d) transmission coefficient against energy for the same. The blue and the red solid lines in each situation represent the $m_s = \pm 1/2$  spin projections, respectively. We have set (a,b) $\lambda= \pi/4$ and (c,d) $\lambda= \pi/2$, the latter showing the extreme localization effect. The values $\epsilon$ and $t$ have been set equal to $0$ and $1$ throughout. For the transmission calculation, the lead, and the lead-system hopping integrals are, $t_\mathrm{lead} = t_\mathrm{lead-system} = 1.5$. $E$ is measured in units of $t$.} 
\label{dosspinhalf}
\end{figure}
These are edge states, and topological in origin, and shown in green and red for  $m_s=\pm 1/2$ in Fig.~\ref{dispersion}(a). They vanish as soon as periodic boundary condition is imposed.
While the first energy $E=0$ is common to all spins, $E=\pm 2$ are special where the bands collapse to lines with width zero. The rest of the values, viz, $E=\pm \sqrt{2}$ are isolated, as we see. As a consequence, at $\lambda=\pi/2$, the entire spectrum for {\it any} half-odd integer spin is composed of just five sharp localized states which speaks for a complete AC-caging of the spins. This is \emph{extreme localization}. We display the phenomenon only for $s=1/2$ here, in Fig.\ref{dosspinhalf}(a) and (c). Others are similar. 
The AC cage topologies for $E=\pm 2$ and for $E=\pm \sqrt{2}$ are much more complicated compared to the pinned flat band at $E=0$. A typical caging for these is displayed in Fig.~\ref{ampli-dist} (a) and (b) for $s=1/2$. The particle can hop only among local clusters marked by blue (for $m_s=1/2$) and red (for $m_s=-1/2$) in the up and down spin projected spaces (red and blue rhombii respectively). The topological edge state at $E=\sqrt{2}$ is clearly seen caged at the left end of the array. The cages are separated from each other by empty circles on which the amplitude of the wave function is zero.  

 For spins $s=1$ and $s=2$, the sharply localized states at $E=\pm 2$ persist, but these are gap-states for the bosonic spin-spectra, flanked on either side by continuous bands. The edge states also appear at other energies (shown by dashed green and red lines in Fig.~\ref{dispersion}(b) for $s=1$). However, the spectra for integer spins, as a whole, never show a complete collapse. Hence, `extreme' localization is ruled out for integer spins, though the topological edge states show up for such spins as well.
\paragraph*{Analyzing the observations} 
A general rotation operator in spin space can be represented as
$
   \hat{\mathcal{R}} = \exp \left[\frac{i \vec{S}\cdot\hat{n} \Delta\phi}{\hbar} \right ]
$ 
which can be re-written, with the definition of the spin matrices via $\vec{S}=(\hbar/2)\vec{\sigma}$, as $\hat{\mathcal{R}}=\exp[i\vec{\sigma}\cdot \hat{n} (\Delta\phi/2)]$. A rotation of a spinor around a single rhombus shown in Fig.~\ref{sample}(a) generates an AC-phase $\Lambda_{AC}$ that makes a spinor $\chi_s$ undergo a change in phase, and transform as $\chi_s \rightarrow \chi_s \exp[i\Lambda_{AC} \vec{\sigma}\cdot \hat{n}]$ from which we can straightforwardly identify the `angle' of rotation as $\Delta\phi=2\Lambda_{AC}$. 
We calculate, for each spin, the product of the individual AC phase factors $\exp[i\lambda\vec{\sigma}\cdot\hat{n}]$ that is implied by the hopping of the particle along the edges of a single rhombus in Fig.\ \ref{sample}. The electric field is taken along the $z$ direction and out of plane, and the sense of rotation has been considered (without loss of generality) clockwise.
Beginning at the left $B$-site of any rhombus and encircling the route $B \rightarrow A \rightarrow B \rightarrow A\rightarrow B$ clockwise. This amounts to an accumulated AC-phase that can be obtained through a sequential product of 
$\exp(-i\lambda\sigma_x)\cdot\exp(-i\lambda\sigma_y)\cdot\exp(+i\lambda\sigma_x)\cdot
\exp(+i\lambda\sigma_y)$. The `effective' change of phase on a full rotation, $\Lambda_{AC}$, involving spin-flip from one spin projection to another as dictated by the Hamiltonian, can then be evaluated through
\begin{equation}
    e^{i \Lambda_{AC} \vec{\sigma}\cdot \hat{n}} = e^{-i\lambda\sigma_x}e^{-i\lambda\sigma_y}
    e^{+i\lambda\sigma_x}e^{+i\lambda\sigma_y}
    \label{effphase}
\end{equation}
and by equating the traces (a gauge invariant quantity) of the matrices appearing on the two sides \cite{avishai-1}. Here, $\vec\sigma$ represents \emph{any} spin, and the matrices accordingly assume dimensions of $(2s+1) \times (2s+1)$. Explicit evaluation of the matrix product of \eqref{effphase} yields for $s=1/2$, $3/2$ and $5/2$, respectively, 
\begin{subequations}
\begin{eqnarray}
    2\cos\Lambda_{AC} & = & F_{1/2}(\lambda), \\
    2~[\cos 3\Lambda_{AC}+\cos \Lambda_{AC}] & = & F_{3/2}(\lambda),  \\
    2~[\cos\Lambda_{AC}+\cos 3\Lambda_{AC} + \cos 5\Lambda_{AC}] & = & F_{5/2}(\lambda),
\end{eqnarray}
\label{halfspintrace}
\end{subequations}
where 
$F_{1/2}(\lambda) = 1-2\sin^4 \lambda$ 
was found already by Avishai and Band~\cite{avishai-1} in an AB caging context. The higher spins are our objects of interest in this paper. We find 
$F_{3/2}=[26+88\cos 2\lambda+17\cos 4\lambda+28\cos 6\lambda-42\cos8 \lambda+12 \cos 10\lambda-\cos 12\lambda]/32$ 
and 
$F_{5/2} = [542+1688 \cos 2\lambda + 622 \cos 4\lambda + 848 \cos 6\lambda - 392 \cos 8\lambda + 16 \cos 10\lambda -621 \cos 12\lambda + 500 \cos 14\lambda -150 \cos 16 \lambda + 20 \cos 18\lambda - \cos 20\lambda]/512$. 
For integer spins, we get for $s=1$ and $s=2$,
\begin{subequations}
\begin{eqnarray}
1+2\cos 2\Lambda_{AC} & = & F_1(\lambda), \\
1+ 2 \cos 2\Lambda_{AC} + 2 \cos 4\lambda_{AC} & = & F_2(\lambda),
\end{eqnarray}
\label{intspintrace}
\end{subequations}
with 
$F_1 = [11 + 8 \cos 2\lambda  + 12 \cos 4\lambda-8 \cos 6\lambda + \cos 8\lambda]/8$, 
and 
$F_2=[211 + 208 \cos 2\lambda +296 \cos 4\lambda - 16 \cos 6\lambda + 44 \cos 8\lambda - 176 \cos 10\lambda + 88 \cos 12\lambda -16 \cos 14\lambda + \cos 16\lambda]/128$. 
Eqs.~\eqref{halfspintrace} and \eqref{intspintrace} immediately reveal that a choice of the Rashba SO coupling $\lambda=\pi/2$ corresponds to $\Lambda_{AC}=\pi$, and hence, to an angle of rotation $\Delta\phi=2 \Lambda_{AC}=2\pi$. For such an `effective rotation' by $2\pi$, the half-odd integer spinors are {\it flipped} after a complete traversal of the loop via an excursion to all the $2s+1$ projections, and the junction between any two consecutive rhombii becomes a {\it node} as a result of a destructive interference. However, the integer  spins retain their phase intact and interfere constructively upon a full rotation. In the former case of half odd-integer spins the amplitudes of the wave functions are {\it caged}, as shown in Fig.~\ref{sample}(a) and in Fig.~\ref{ampli-dist}. Such a distribution is effectively {\it decoupled} from the next cluster of non-zero amplitudes by a set of sites where the amplitudes are zero. \footnote{If we released a spin-half particle at one of the A sites, then on a full rotation the destructive interference would occur at A-vertices. Non-zero amplitude in this case would be pinned at the B-vertices.} This explains the interference mechanism for the spin-selective extreme localization.
\begin{figure}[tb]
(a)\includegraphics[width=.4\columnwidth]{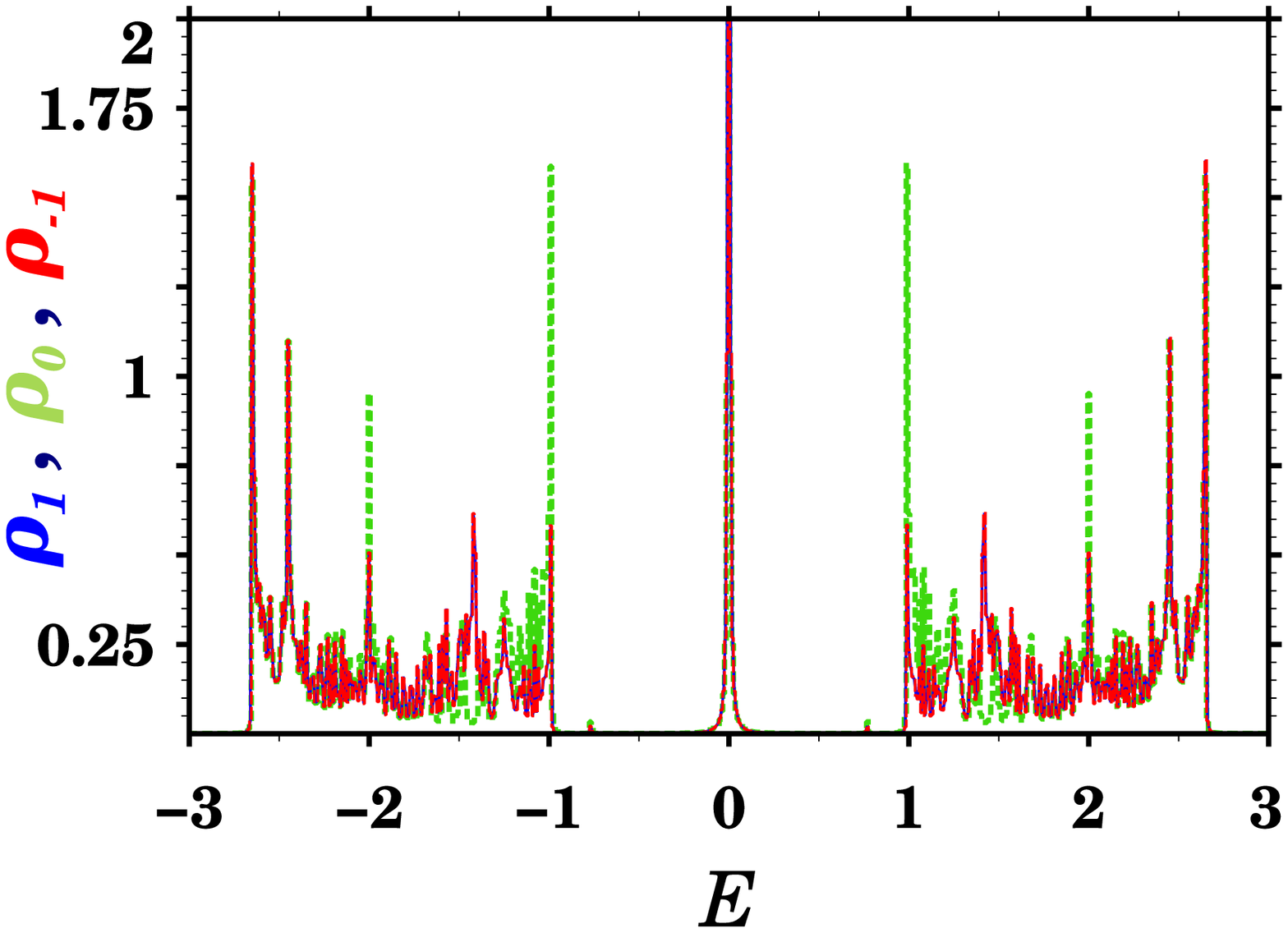}
(c)\includegraphics[width=.4\columnwidth]{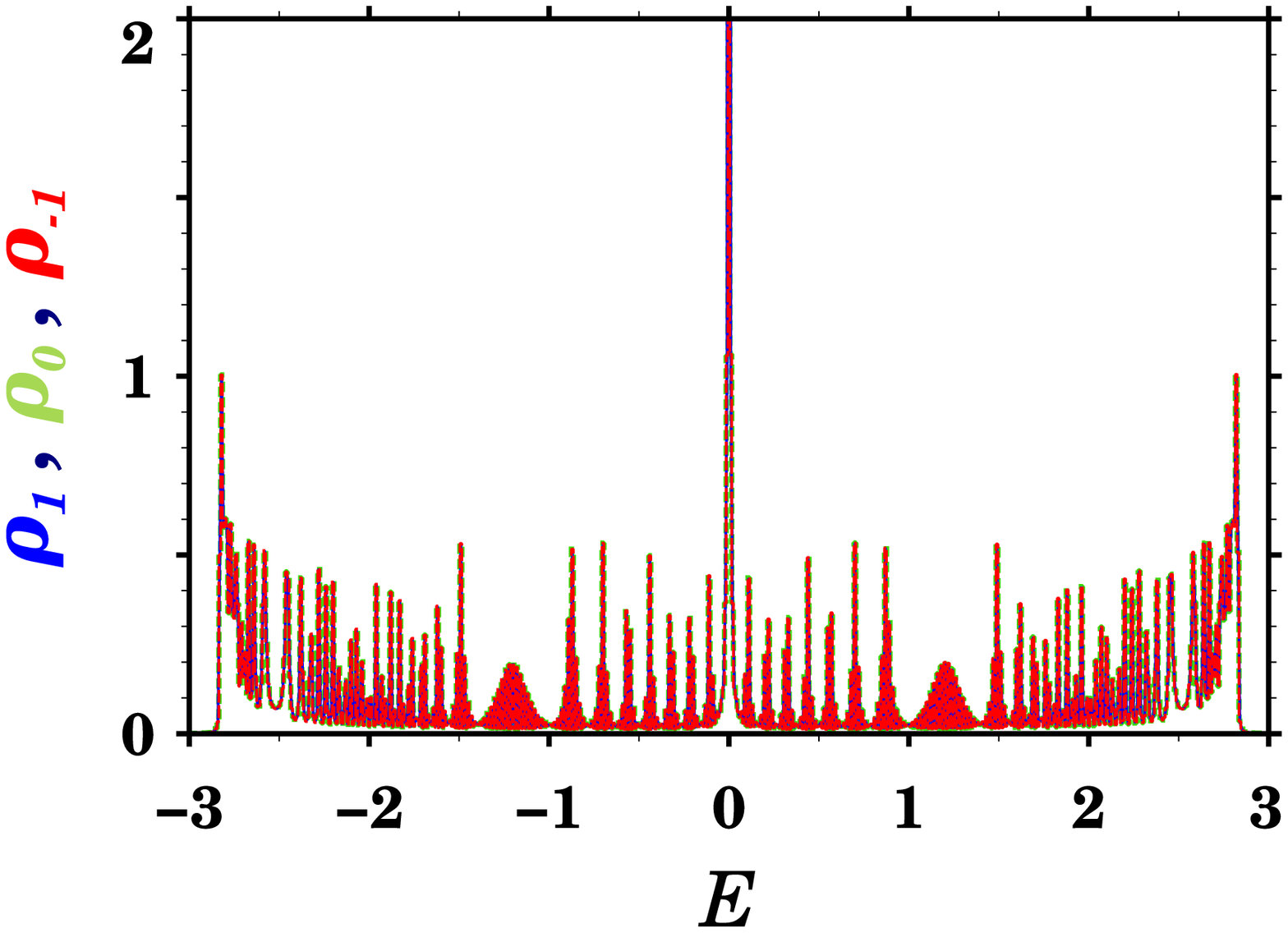}\\

(b)\includegraphics[width=.4\columnwidth]{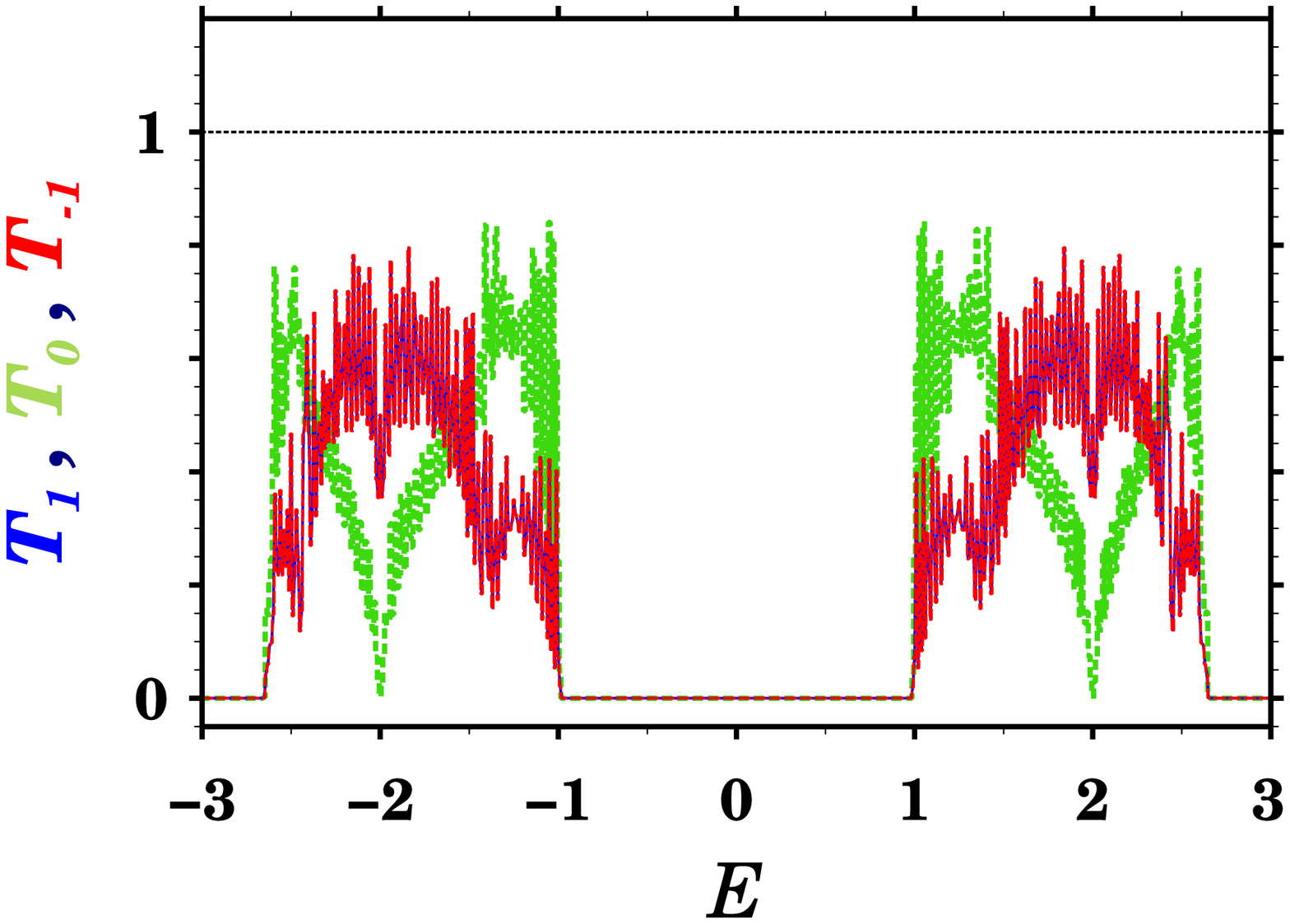}
(d)\includegraphics[width=.4\columnwidth]{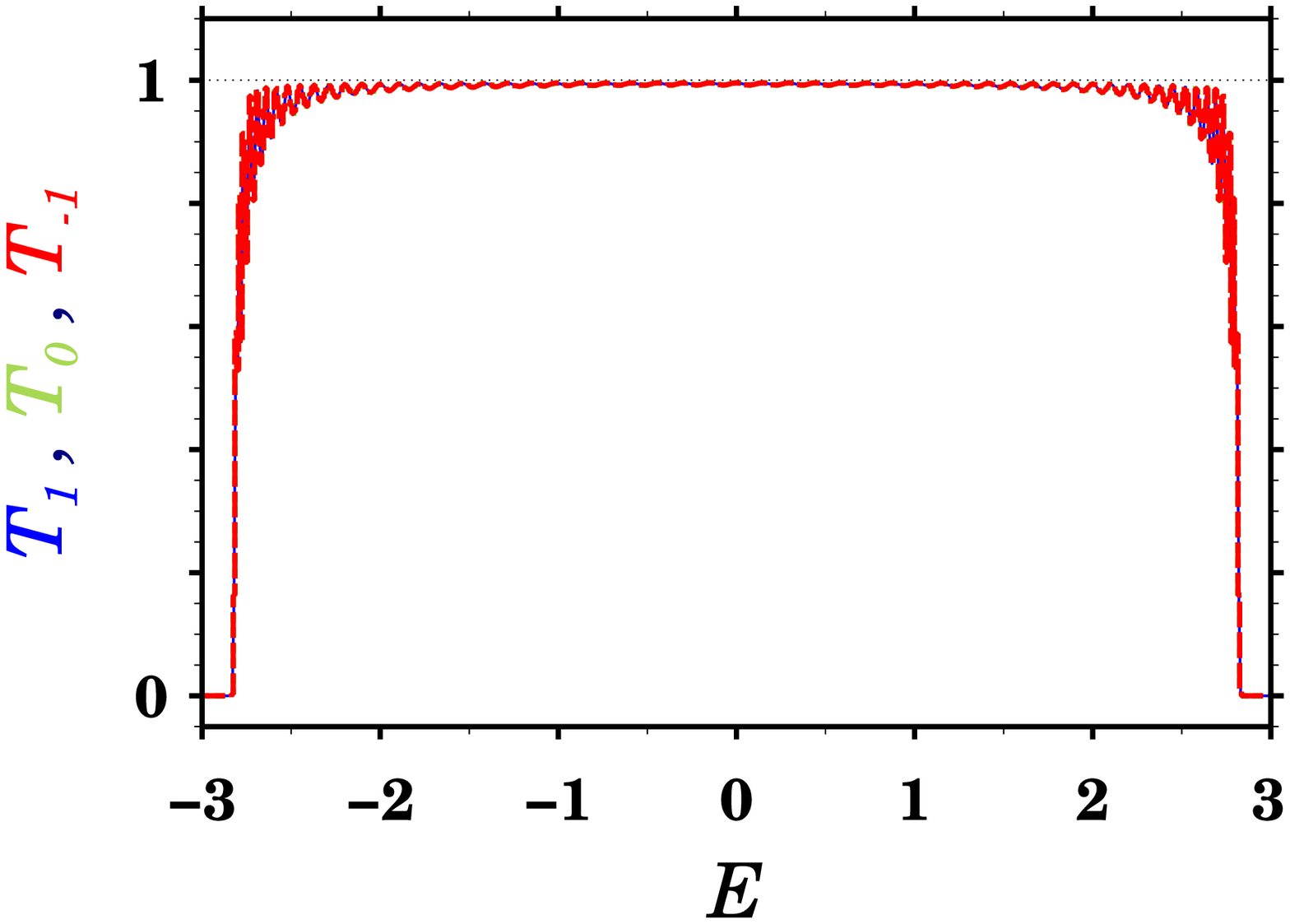}
\caption{
(a,c) DOS and (b,d) transmission coefficient (for $30$ rhombii) against energy for $s=1$. The DOS for the $m_s = \pm 1 $ states are identical and hence we only show $m_s=-1$ (red). The same holds for the transport with $m_s = \pm 1 $. The green color represents the $m_s=0$ projection, where a clean spin-selective transmission, making the $m_s=0$ case differ from $m_s=\pm 1$, is observed. As in Fig.~\ref{dosspinhalf}, $\lambda=\pi/4$ and $\pi/2$ in (a,b) and (c,d), respectively. The values of $\epsilon$, $t$, $t_\mathrm{lead}$ and $t_\mathrm{lead-system}$ are also as in Fig.\ \ref{dosspinhalf}.}
\label{dosspinone}
\end{figure}

An intuitively appealing check to the above observation is provided by a real space renormalization group (RG) scheme that decimates the vertices $A$ using Eq.~\eqref{difference}. The array of rhombii gets reduced to an effective linear chain of pseudo-atomic sites (with renormalized potentials) residing at locations $B$ only (Fig.~\ref{sample}). The renormalized hopping integral connecting the consecutive $B$ sites on this {\it effectively} one dimensional array (diagonally opposite $B$ sites in the original array) is given by,
\begin{equation}
   \tilde{\mathbf{t}}_{BB} = t~e^{i\lambda\sigma_x} (E - \epsilon)^{-1} t~e^{i\lambda\sigma_y} + t~e^{i\lambda\sigma_y} (E - \epsilon)^{-1} t~e^{i\lambda\sigma_x} .
   \label{trenorm}
\end{equation}
For $s=1/2$, Eq.~\eqref{trenorm} becomes 
\begin{equation}
   \tilde{\mathbf{t}}_{BB} = 
     \begin{pmatrix}
  \frac{2t^2 \cos^2\lambda}{(E-\epsilon)} &   \frac{(2+2i)t^2 \cos\lambda~\sin\lambda}{(E-\epsilon)}  \\ 
  -\frac{(2-2i)t^2 \cos\lambda~\sin\lambda}{(E-\epsilon)} & \frac{2t^2 \cos^2\lambda}{(E-\epsilon)}  
     \end{pmatrix}.
   \label{trenormspinhalf}
\end{equation}
It is clear from \eqref{trenormspinhalf} that a choice of $\lambda=\pi/2$ renders $\tilde{t}_{BB}$ into a null matrix. This indicates a complete `cut-off' between the pair of sites occupying the $B$ positions in a rhombus, prohibiting any propagation in the longitudinal direction. The transmission coefficient across the array naturally becomes zero. This happens for all the half odd integer spins and is not seen for an integer spin. However, for certain localized `gap states' the hopping integral for integer spin states doesn't become zero immediately, but eventually flows to zero after a finite number of RG iterations - a common signature of localization. Still, the phenomenon of a complete collapse of the spectrum into an extreme localization picture does never happen for them.

\paragraph*{Transmission coefficient and its spin selectivity} 
The transmission coefficient for a particle entering into the system at a spin state $\sigma$ and ejecting out at a state $\sigma'$ is calculated following the standard procedure as outlined for example, by Datta et al.~\cite{datta1990}, and is given by 
$
\mathcal{T}_{\sigma\sigma'} = \mathrm{Tr}~\left [\Gamma^i_{\sigma}~\mathcal{G}_\mathrm{comp,r}\Gamma^j_{\sigma'}~\mathcal{G}_\mathrm{comp,a} \right ]
$.
Here $\Gamma^{i,j}$ denote the matrices that connect the system to the leads, and $\mathcal{G}_\mathrm{comp}^r$ and $\mathcal{G}_\mathrm{comp}^a$ are the retarded and advanced Green's functions of the lead-system-lead composite system, obtained 
following L\"{o}wdin's partition technique \cite{lowdin}. The general formula is 
$
 \mathcal{G}_\mathrm{comp} = \left (E- \mathcal{H} -\Sigma^{L}_\sigma - \Sigma^{R}_\sigma \right )^{-1} 
$, where $E$ is the energy of the incoming particle and $\Sigma^{L(R)}_\sigma$ is the self-energy contribution of the leads, chosen appropriately to describe retarded/advanced cases. 
Of particular interest is the total collapse of the extended spectrum, as obtained for say, $\lambda=\pi/4$ in the spin half case, into an extremely localized set of just five spikes when $\lambda=\pi/2$. In this case the rhombic array turns out to be totally opaque to the incoming spin $s=1/2$, as seen in Fig.~\ref{dosspinhalf}(d). This has been checked to be also true for the half odd integer spins $3/2$ and $5/2$, and we believe it to hold in general. 
This is not the case for integer spins. For $s=1$ and $\lambda=\pi/4$, the transport of $m_s=\pm 1$ dominates the $m_s=0$ state, say, at $E=\pm 2$, while there are other regimes of the Fermi energy where the $m_s=0$ transport channel dominates the other spin channels. This provides an example of what one may call a {\it spin de-multiplexer}. For $\lambda=\pi/2$ and $s=1$ (and for all other integral spins as well), the network decouples into $2s+1$ independent rhombic arrays, each array representing a perfectly periodic system connected to clean leads. This can be easily verified from Eq.~\eqref{difference} explicitly in terms of the appropriate hopping matrices. The transport becomes unattenuated and identical for $m_s=0$ and $m_s=\pm 1$, as is displayed in Fig.~\ref{dosspinone}(d).

\paragraph*{Conclusions} 


Our model is a spin-resolved version of a photonic AB cage recently demonstrated \cite{alex}. We find an analogous AC caging effect, however, the effect being retained only for half-odd integer spins while integer spins do not show caging. This leads to a dramatic difference in the localization and transport characteristics where only the half-odd integer spins can be chosen to transport at selected and SO strength-tunable energies, while integer spins have wider transmission windows with selectable spin-projections. 
While UC atomic gases, with systems of higher spins now routinely studied, appear as obvious examples where to realize our proposal, we also think that solid-state devices of, say, coupled quantum dots appear as promising candidates.

\paragraph*{Acknowledgements}
We thank D.\ Leykam for sharing his results on helical lattices prior to publication. This work has been supported jointly by the UGC, India and the
British Council through UKIERI, Phase III, reference numbers F.\ 184-14/2017(IC) 
and UKIERI 2016-17-004 in India and the U.K., respectively. A.C.\ gratefully acknowledges research grant FRPDF of Presidency University and an IAS residential fellowship at University of Warwick, where this work was completed. A.M.\ is thankful to DST, India, for awarding her the INSPIRE fellowship IF160437. 
UK research data statement: all data are directly available within the publication.

\bibliographystyle{apsrev4-1}
%

\clearpage

\end{document}